\documentclass[11pt]{article}
\usepackage{amsmath}
\usepackage{graphicx}
\usepackage{caption}
\usepackage{authblk}
\usepackage{mathpazo} % Palatino font
\usepackage[a4paper,margin=1in]{geometry}
\usepackage[numbers,sort&compress]{natbib} % For bibliography
\usepackage{setspace}
\usepackage{siunitx}
\usepackage{hyperref}
\usepackage[pagewise]{lineno}  % or just \usepackage{lineno}

\newcommand{\affilfont}{\small} % Affiliation font
\newcommand{\authorfont}{\normalsize} % Author font
\newcommand{\abstractfont}{\normalsize} % Abstract font
\newcommand{\titlefont}{\large} % Title font
\newcommand{\maintextfont}{\normalsize} % Main text font
\usepackage{microtype}

\title{\titlefont \textbf{Cavity-Driven Attractive Interactions in Quantum Materials}}
\author[1]{\authorfont F. Helmrich}
\author[1,2]{\authorfont H. S. Adlong}
\author[1]{\authorfont I. Khanonkin}
\author[1]{\authorfont M. Kroner}
\author[1]{\authorfont G. Scalari}
\author[1]{\authorfont J. Faist}
\author[1]{\authorfont A. \.{I}mamo\u{g}lu}
\author[1]{\authorfont T. F. Nova}
\affil[1]{\affilfont Institute for Quantum Electronics, ETH Zürich, Auguste-Piccard-Hof 1, 8093 Zürich, Switzerland}
\affil[2]{\affilfont Institute of Theoretical Physics, ETH Zürich, Wolfgang-Pauli-Strasse 27, 8093 Zürich, Switzerland}
\date{}

\begin{document}
% \linenumbers
  
\maketitle

\begin{abstract}
\abstractfont
\textbf{Many-body phenomena in quantum materials emerge from the interplay among a broad continuum of electronic states, and controlling these interactions is critical for engineering novel phases. One promising approach exploits fluctuations of the vacuum electromagnetic field confined within optical cavities to tailor electronic properties \cite{schlawin2022,garciavidal2021}. Here, we demonstrate that cavity photons can mediate attractive interactions in a tunable van der Waals material and reorganize a continuum of electron–hole transitions into an exciton-like state. We introduce a broadband, sub-wavelength time-domain microscope that integrates exfoliated, dual-gated two-dimensional quantum materials into a terahertz cavity. This approach enables the first-ever measurement of the field-tunable bandgap of bilayer graphene \cite{Zhang2009,mccann2013} at terahertz frequencies \cite{ju2017} while revealing ultrastrong coupling \cite{ciuti_quantum_2005,FriskKockum2019,forndiaz2019} with a vacuum Rabi frequency exceeding $\Omega_{Rabi}/\omega\approx 40\%$ of the bare photon energy. Crucially, we identify a novel cavity-induced resonance emerging from the interband continuum that resembles Coulomb-bound excitons and remains stable across a broad temperature range. By uniting longstanding theoretical predictions with advanced experimental techniques, our findings open new avenues for designing and probing unique light–matter states and realizing hybrid correlated phases in quantum materials.}
\end{abstract}
\maintextfont
The control of quantum materials through light has developed into a significant research direction in materials science and condensed matter physics \cite{delatorre2021,disa_engineering_2021}. At the forefront of modern research is the vision that ground state properties of materials could be modified through interaction with cavity-confined electromagnetic fields \cite{schlawin2022,garciavidal2021}. Even in the absence of external stimuli, matter can couple to quantum fluctuations of the vacuum electromagnetic field, leading to the formation of hybrid light-matter states known as cavity polaritons \cite{basov2021}. In the ultrastrong coupling regime—where the interaction strength rivals the energy of material excitations—these novel hybrid phases are predicted to exhibit markedly altered macroscopic properties, with ground states that contain virtual light and matter excitations \cite{ciuti_quantum_2005,FriskKockum2019,forndiaz2019}. Theoretical work has proposed that such coupling may induce superconductivity \cite{schlawin2019,lu_cavity-enhanced_2024}, trigger ferroelectric phase transitions \cite{ashida2020,latini2021}, stabilize superradiant excitonic insulators \cite{debernardis2018,mazza2019,andolina2019,andolina2020}, or control intertwined orders \cite{li2020}. Experimental studies further underscore the impact of confined fields: in polariton chemistry, cavity coupling to molecular vibrations modulates chemical reactivity \cite{thomas2016}, while in solid-state systems, cavity fields have tuned metal–insulator transitions \cite{Jarc2023} and non-perturbatively altered transport in both integer and fractional quantum Hall phases \cite{appugliese2022,enkner2024enhancedfractionalquantumhall,graziotto2025_stripes}. Yet, clear experimental evidence demonstrating the use of cavities to modify, induce, or control quantum phases in condensed matter remains scarce.\\
\begin{figure}[h!]
    \centering
    \includegraphics[width=\textwidth]{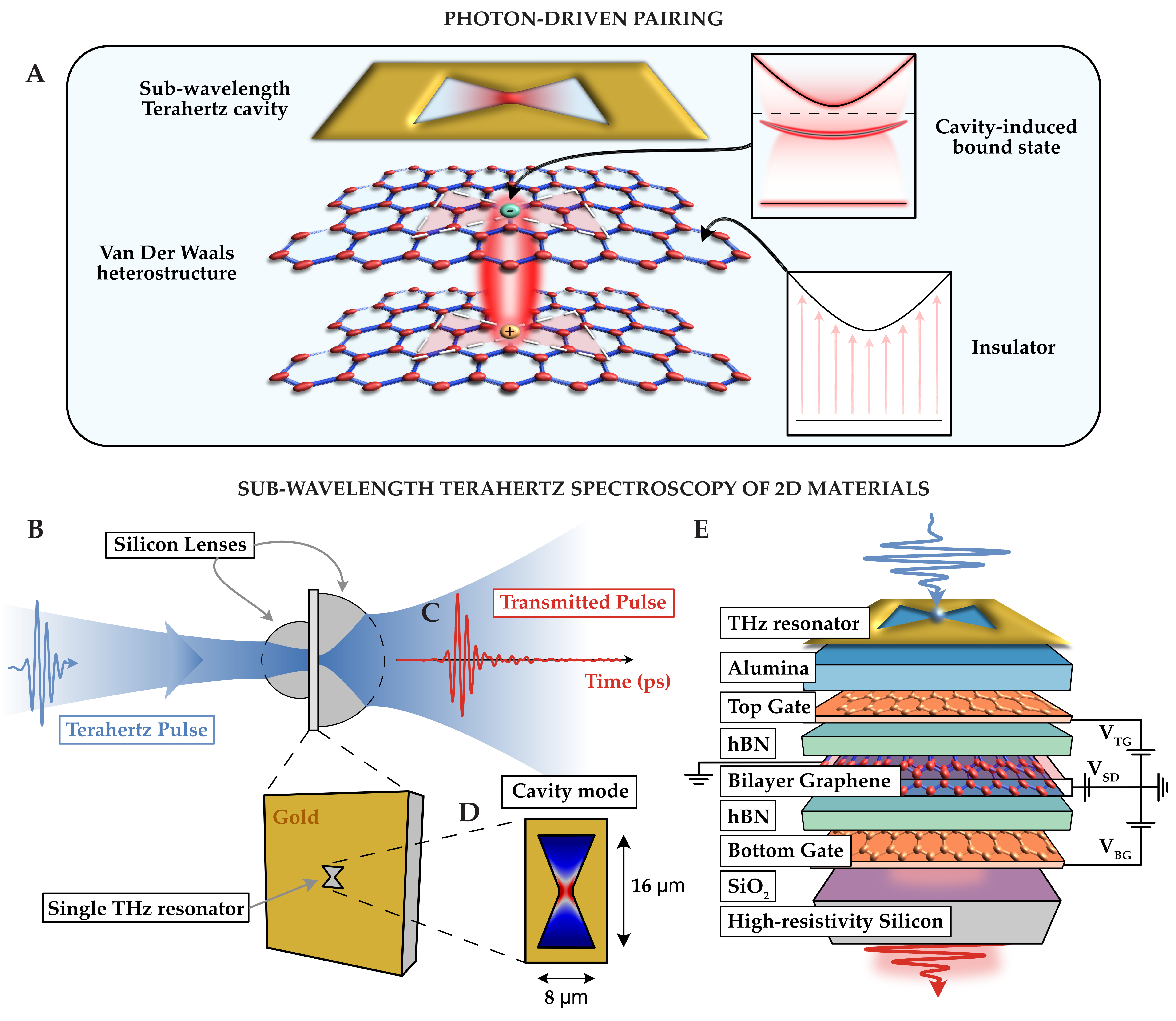}
    \caption{\textbf{Vision and Setup |} \textit{Cavity-induced bound states :} \textbf{A} A two-dimensional insulating material is coupled to a sub-wavelength terahertz cavity. In the insulator, a broad continuum of interband electron–hole transitions—from the valence band to the conduction band—underpins its optical response. When the cavity is tuned to resonance with the material’s bandgap, the localized cavity field acts as an effective pairing interaction that binds electron–hole pairs across a wide momentum range, reorganizing the continuum into a discrete, bound state. This cavity-induced excitonic state exemplifies how strong light–matter interaction can be harnessed to engineer novel many-body phases in gate-tunable 2D systems. \textit{Sub-wavelength terahertz spectroscopy of  bilayer graphene :}\textbf{B} A few-cycle terahertz pulse (blue line) is focused on a 2D heterostructure with a bow-tie resonator and solid immersion lenses. \textbf{C} Terahertz pulse transmitted through lens-resonator assembly (red line) as detected by electro-optic sampling. \textbf{D} In-plane electric field of the resonant cavity mode calculated with finite-element methods. The field is concentrated on the apex of the bow-tie. \textbf{E} Lateral view of the sample structure. The heterostructure consists of Bernal bilayer graphene, few-layer graphite gates and hbN dielectrics placed on a high-resistivity Si/SiO$_2$ substrate. The heterostructures is insulated from the bow-tie resonator using alumina dielectric.}
    \label{fig:image1}
\end{figure}
Here, drawing on predictions \cite{khurgin2001,Cortese_2019} and experiments \cite{cortese_excitons_2021,bloch_giant_1998} in quantum well systems, we aim to demonstrate that vacuum electromagnetic fields confined within a terahertz cavity can mediate an effective attractive interaction among charged particles and drive the emergence of a bound state from a continuum of electron--hole transitions, leading to a profound reshaping of the excitation spectrum in a van der Waals (VdW) material (Fig. \ref{fig:image1}A).\\
To address this goal, we have developed an experimental platform specifically designed to overcome the limitations that have hindered progress in exploring light-matter coupling and spectroscopy in micro-structured quantum materials. Our strategy focuses on VdW materials, which are uniquely advantageous due to their naturally enhanced light-matter coupling—thanks to reduced dimensionality—and their rich variety of quantum and correlated phases, ranging from superconductivity \cite{Cao2018} and Wigner-Mott states \cite{Regan2020,Tang2020,shimazaki_strongly_2020} to exotic magnetism \cite{Ciorciaro2023}. Crucially, multiple phases can often be accessed within a single device through electrostatic gating, making them particularly suitable for investigating light-matter interactions across a broad range of many-body phenomena.\\
A common strategy to realize ultrastrong coupling is to boost the light–matter interaction by enhancing the vacuum electromagnetic field, which scales as ${E_{vac,rms}}\propto\sqrt{\frac{\omega_c}{V_c}}$, where $\omega_c$ is the cavity frequency and $V_c$ is the cavity mode volume \cite{Kavokin_Baumberg_2017}. This approach requires confining the electromagnetic field within a cavity mode that is significantly smaller than the photon wavelength.
However, integrating two-dimensional (2D) materials into such cavities poses substantial technical challenges, especially since ultrastrong light–matter coupling typically occurs near collective excitations in the terahertz range. While sub-wavelength localization of terahertz fields can be efficiently achieved using micro-structured metallic resonators, conventional approaches rely on metamaterials of repeating cavities placed atop the material of interest \cite{scalari2012,Jin_Razzari_ADV_FUNC_2025}. This method is incompatible with exfoliated 2D materials, which are usually limited to in-plane dimensions of just a few microns. \footnote{Large-area 2D materials obtained by direct growth \cite{Valmorra2013,Ren2012} or exfoliation using metallic layers \cite{Handa_2024,Jin_Razzari_ADV_FUNC_2025} are better adapted to the sample-size requirements of terahertz spectroscopy, but are not commonly assembled into tunable high-quality heterostructures achievable with exfoliation \cite{wang2013}.}
Recent advancements have enabled spectroscopy of individual metallic resonators by adapting solid immersion lenses to the terahertz regime \cite{rajabali2022}. Building on this concept, we have developed a method that simultaneously delivers the broadband frequency operation essential for quantum materials spectroscopy \cite{Basov2011} and strong field confinement, all while maintaining compatibility with gate-tunable van der Waals devices.
% \begin{figure}[h!]
%     \centering
%     \includegraphics[width=\textwidth]{Images/Figure_1.3.png}
%     \caption{\textbf{Sub-wavelength terahertz spectroscopy of cavity-coupled bilayer graphene 
%     |} \textbf{A} A few-cycle terahertz pulse (blue line) is focused on a 2D heterostructure with a bow-tie resonator and solid immersion lenses. \textbf{B} Terahertz pulse transmitted through lens-resonator assembly (red line) as detected by electro-optic sampling. \textbf{C} In-plane electric field of the resonant cavity mode calculated with finite-element methods. The field is concentrated on the apex of the bow-tie. \textbf{D} Lateral view of the sample structure. The heterostructure consists of Bernal bilayer graphene, few-layer graphite gates and hbN dielectrics placed on a high-resistivity Si/SiO$_2$ substrate. The heterostructures is insulated from the bow-tie resonator with alumina dielectric.}
%     \label{fig:image1}
% \end{figure}

\section*{Experimental Setup and Material System}
In our setup (Fig. \ref{fig:image1}B), a free-space terahertz beam is tightly focused using a high-refractive-index, transparent hyper-hemispherical silicon lens. The beam impinges on a high-resistivity silicon substrate coated with gold, except for a small complementary bow-tie resonator designed to function over a broadband range of terahertz frequencies. This resonator steers the field onto a spot smaller than 1.5 microns, as represented by the antenna mode in Fig. \ref{fig:image1}D. The transmitted terahertz field (Fig. \ref{fig:image1}C), with a bandwidth ranging from 1 to 6 THz (4 to 24 meV), is then re-collimated by a secondary hemispherical silicon lens and detected using conventional far-field methods, such as electro-optic sampling within a nonlinear crystal. Crucially, only the field that traverses the micron-sized resonator focus is transmitted, enabling sub-wavelength, broadband terahertz spectroscopy of micron-sized samples positioned beneath the antenna focus (Fig. \ref{fig:image1}E). The cavity gap size is chosen to maximize field confinement and spatial resolution, while effectively mitigating nonlocal effects \cite{rajabali_polaritonic_2021}.\\

We chose Bernal-stacked bilayer graphene (BLG) as a material system due to its ideal characteristics for studying light-matter interactions: while the material is intrinsically a semimetal with quasi-parabolic bands at the $K$ and $K'$ points, a unique "Mexican hat" band gap can be opened and controlled by the application of a displacement field perpendicular to the graphene layers \cite{mccann2013}. As a consequence, BLG offers strong optical excitations such as band-to-band transitions \cite{Zhang2009} and excitons \cite{ju2017} that can be continuously tuned from the mid-infrared to the terahertz range, perfectly aligning with our setup capabilities.

\begin{figure}[h!]
    \centering
    \includegraphics[width=\textwidth]{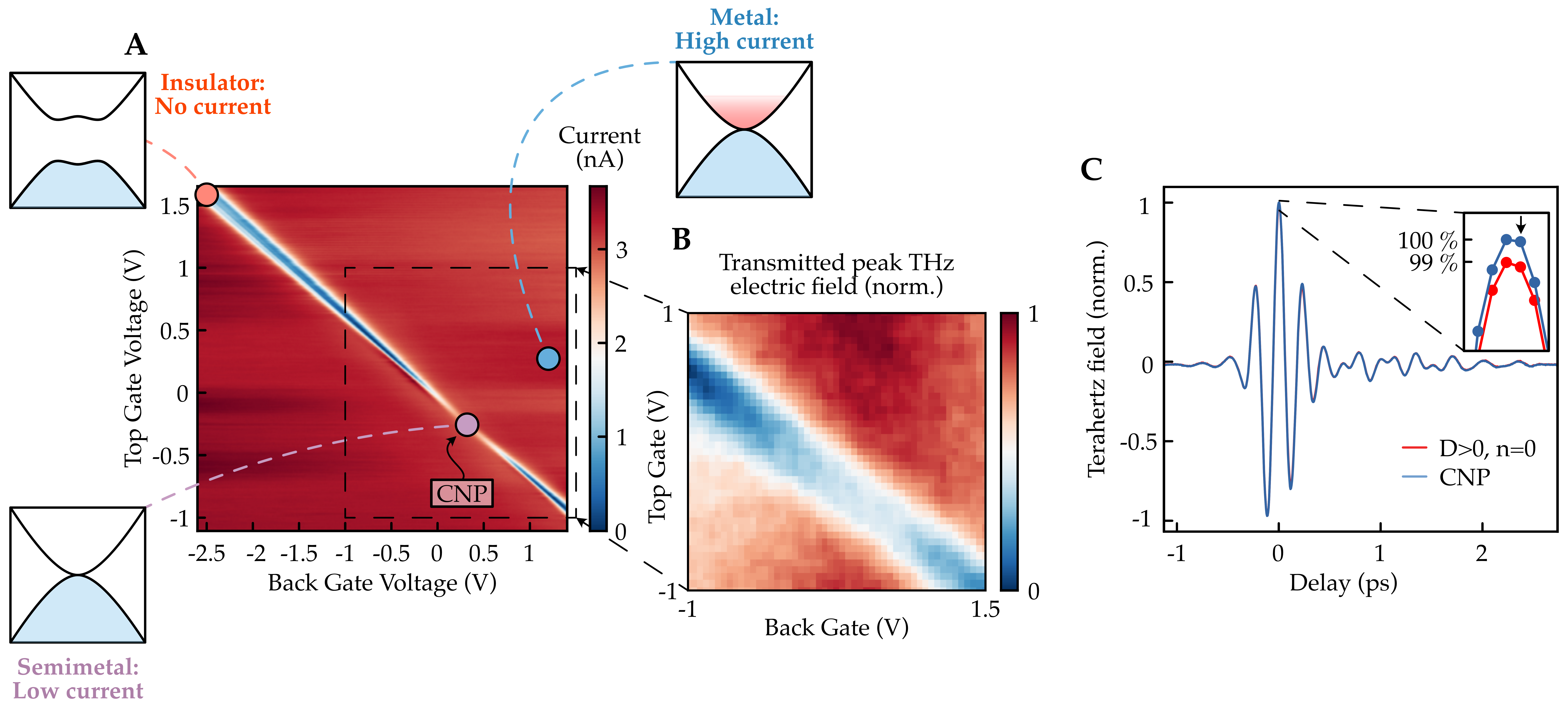}
    \caption{\textbf{Transport characteristics and terahertz transmission of the bilayer graphene device.} \textbf{A} Dual-gate-dependent 2-point transport measurement of BLG. The colored dots indicate key points: the charge neutrality point (CNP, purple), the insulating phase where the THz reference spectrum is recorded (red, at \( D = \SI{-0.38}{V/nm} \)), and a representative metallic state (blue). \textbf{B} Terahertz field transmission at a fixed time delay, with varying top and back gate voltages. The chosen time delay corresponds to the THz field maximum (indicated by the black arrow in panel C inset). The data has been normalized to lie within the range [0, 1]. \textbf{C} Transmitted terahertz pulses when BLG is at the CNP (blue) and in an insulating state with $D>0$ (red). The inset shows a closer view of the maximum of the waveforms. Both traces are normalized to the maximum of the blue curve. All measurements were taken at \SI{5}{K}.}
    \label{fig:image2}
\end{figure}

The device (see Fig. \ref{fig:image1}E) consists of BLG encapsulated between two insulating hexagonal boron nitride (hBN) layers and is fully tunable with two few-layer graphite gates, allowing independent control of the charge density $n = \frac{\epsilon_{\text{hBN}} \epsilon_0}{e} \left( \frac{V_\text{TG}}{d_\text{TG}} + \frac{V_\text{BG}}{d_\text{BG}} \right)$ and the out-of-plane displacement field $D = \frac{\epsilon_{\text{hBN}} \epsilon_0}{2} \left( \frac{V_\text{TG}}{d_\text{TG}} - \frac{V_{\text{BG}}}{d_\text{BG}} \right)$. Here, $V_\text{TG (BG)}$\footnote{Note that, for simplicity, we include the shift of charge neutrality from \SI{0}{V} in the definition of $V_\text{BG}$ and $V_\text{TG}$ (see supplementary material for details).}, $d_\text{TG (BG)}$, $e$, $\epsilon_{\text{hBN}}$, and $\epsilon_0$ are the top and bottom gate voltages, the correspoding hBN layer thicknesses, the elementary charge, the hBN dielectric constant, and vacuum permittivity, respectively. Two Cr/Au etch contacts to BLG \cite{wang2013} allow us to probe the transport properties of the device. The stack is completed by the terahertz resonator, which is electrically insulated from the heterostructure and its electrical connections via a layer of alumina grown by atomic layer deposition.
% (8 nm Ti/192 nm Au) I would suggest to bring to the supplementary
% ($\approx3.4$ \cite{Pierret2022}) also
% We chose few-layer graphite as gate material to minimize THz absorption by the gates.

\section*{Transport Characteristics and Terahertz Transmission}
We first characterized the gate-voltage dependent two-point conductance of our device at \SI{5}{K} (Fig. \ref{fig:image2}A), identifying the typical transport regimes of bilayer graphene. For $V_\text{TG} = -V_\text{BG}\frac{d_\text{TG}}{d_\text{BG}}$, the out-of-plane displacement field $D$ breaks inversion symmetry between the two graphene layers, opening a band gap that results in a minimum in conductance typical of an insulating state. This line corresponds to the diagonal charge neutrality region in Fig. \ref{fig:image2}A, where $n=0$. For $V_{TG} = V_\text{BG}\frac{d_\text{TG}}{d_\text{BG}}$, we increase the carrier concentration without opening a band gap, thus making the material metallic and increasing the conductivity. At the Charge Neutrality Point (CNP), where both $n=0$ and $D=0$, the material behaves like a semimetal (purple dot in Fig. \ref{fig:image2}A).\\
\begin{figure}[h!]
    \centering
    \includegraphics[width=\textwidth]{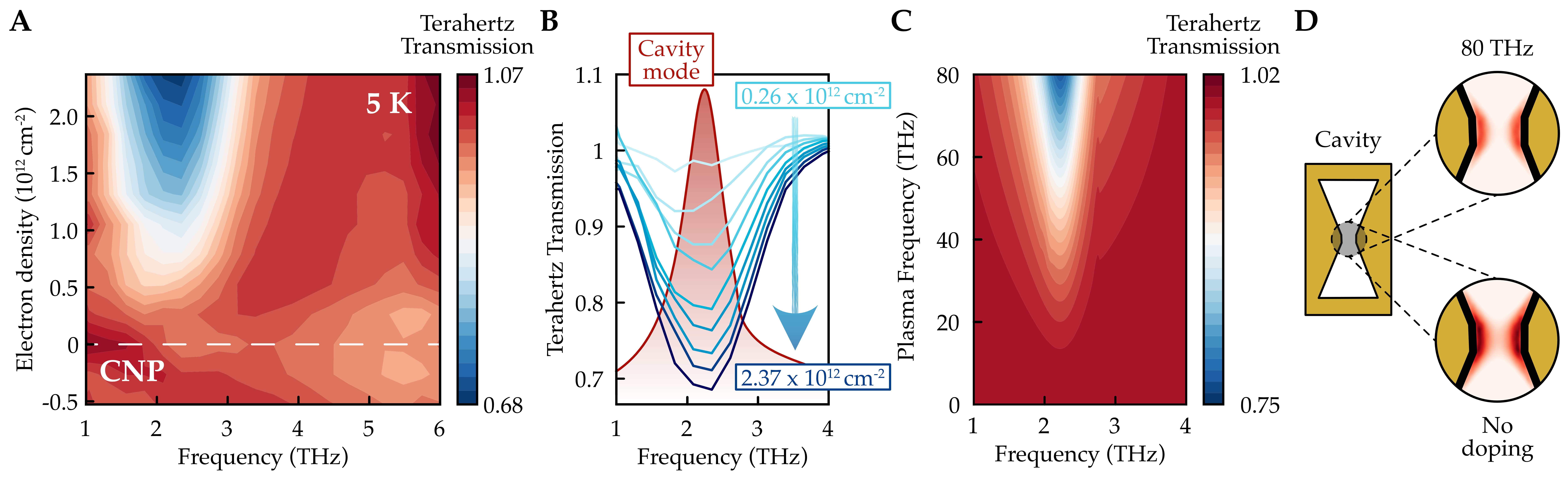}
    \caption{\textbf{Cavity modulation by electron doping.} \textbf{A} Contour plot of terahertz transmission spectra of BLG as a function of electron density $n$ at zero displacement field ($D=0$). All spectra are normalized to the insulating state at $D=\SI{-0.38}{V/nm}$. The measurements were taken at \SI{5}{K}. \textbf{B} Line cuts of the spectra in panel A showing the dip in transmission. The red line represents the bare cavity transmission spectrum, calculated using finite-element methods. \textbf{C} Finite-element simulations of the transmission spectra of the bow-tie resonator coupled to a conductive 2D material modeled by a Drude conductivity with variable plasma frequency. The spectra are referenced to the transmission spectrum of the bare cavity. \textbf{D} Simulated electric-field mode profiles for two of the spectra of panel C. The mode profiles are evaluated at the cavity resonance for a plasma frequency of \SI{0}{THz} and \SI{80}{THz}.}
    \label{fig:image3}
\end{figure}

Figure \ref{fig:image2}C shows the terahertz pulse transmitted through the resonator, material stack, and silicon lenses when the device is biased at the CNP (blue line). As the system deviates from the CNP by varying $n$ and $D$, the terahertz amplitude changes by a few percent. This trend is illustrated by the transmission data in the insulating phase ($D \neq 0$ and $n=0$) as shown by the red line in the same figure.

Figure \ref{fig:image2}B presents the gate-voltage dependence of the terahertz transmission measured at a fixed time delay (indicated by the black arrow in the inset of Fig.~\ref{fig:image2}C). The transmission data closely resembles the transport measurements (Fig.~\ref{fig:image2}A), with a distinct charge neutrality line that mirrors the insulating region observed in transport—exhibiting nearly identical slope and intercept, though with a significantly larger width.\footnote{Note that the single-time delay measurement does not provide any spectral information, which can only be obtained by measuring the full time-domain trace for each ($V_{BG}$, $V_{TG}$) pair.} This result clearly demonstrates that the terahertz signal is sensitive to the electronic state of bilayer graphene (BLG) while remaining unaffected by other elements of the stack, including the charged gates. We attribute this insensitivity to the reduced charge mobility and the consequent suppression of the Drude-like response in the non-encapsulated gates relative to BLG.
\section*{Doping-induced cavity mode screening}
To gain insight into the optical properties of the coupled BLG/cavity system, we measured the spectrally resolved terahertz transmission of doped BLG. Figure \ref{fig:image3}A shows the terahertz spectra measured for different charge densities with $n\neq0$ and $D = 0$. Each spectrum is referenced to the transmission taken for BLG in an insulating state ($D=\SI{-0.38}{V/nm}$ and $n=0$; red dot in Fig.  \ref{fig:image2}A), where the band gap exceeds the terahertz bandwidth of our setup and the material becomes transparent for THz radiation. Effectively, our data are normalized to the transmission of the bare cavity.
As the electron density increases, a pronounced dip in the terahertz transmission emerges around \SI{2.3}{THz}. The dip coincides with the peak position of the simulated cavity transmission spectrum (Fig. \ref{fig:image3}B), consistent with observations in ref.~\cite{Valmorra2013}. As the resonance frequency and linewidth remain mostly unchanged (line cuts in Fig.~\ref{fig:image3}B), we conclude that the primary effect is the screening of the cavity mode by the injected electrons, which reduces the cavity transmission.  The coupled system acts as a terahertz attenuator with modulation depths of around 30\% in power for relatively low doping.
We performed finite-element simulations modeling the electromagnetic interaction between our bow-tie resonator (Fig.~\ref{fig:image1}C) and a conductive two-dimensional material described by a Drude model. By varying the carrier density, the simulations reproduce the experimentally observed transmission dip at the cavity resonance (Fig.~\ref{fig:image3}C) and reveal a reduced mode intensity at higher carrier densities (Fig.~\ref{fig:image3}D), corroborating the role of carrier-induced mode screening.

\begin{figure}[h!]
    \centering
    \includegraphics[width=\textwidth]{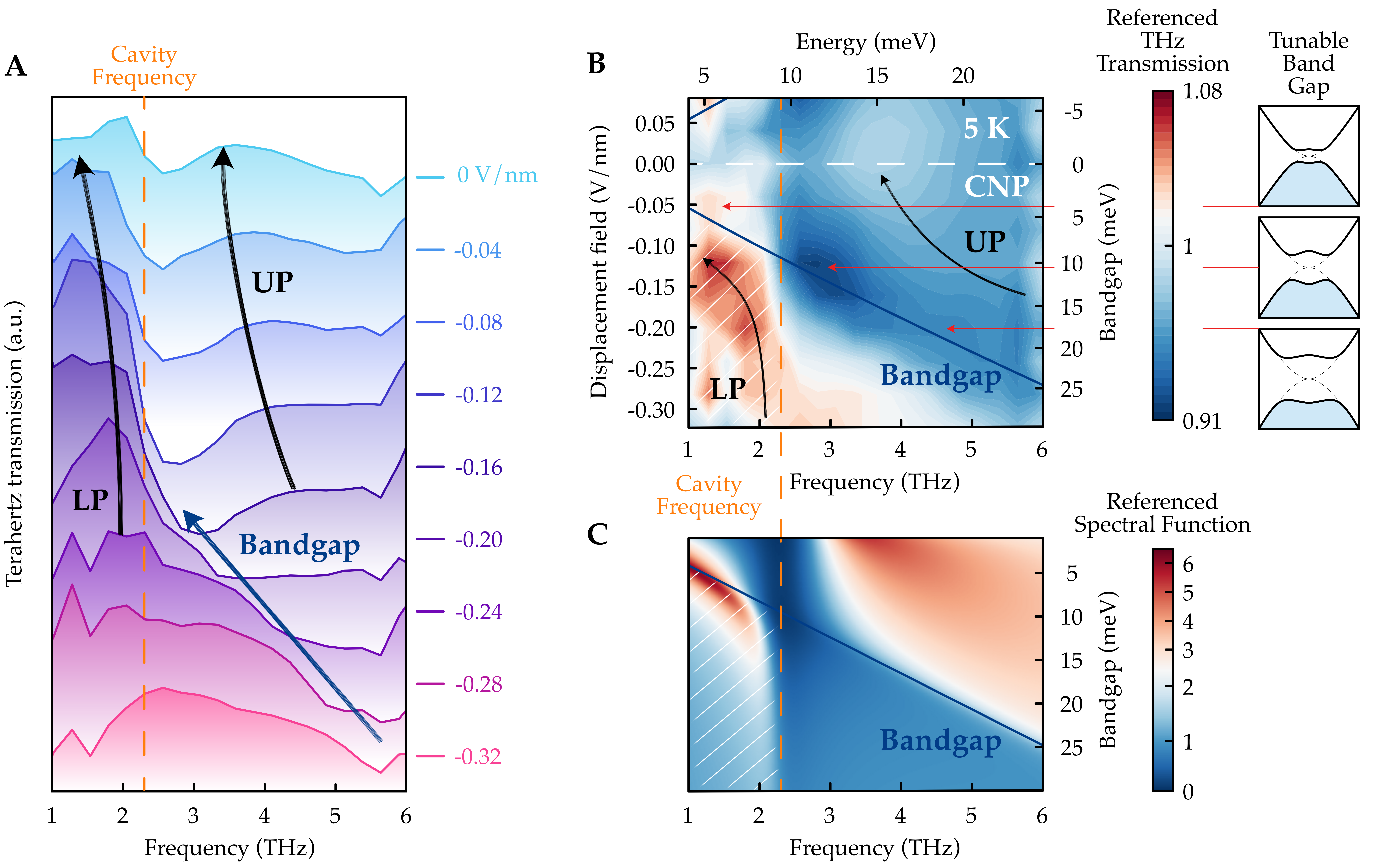}
    \caption{\textbf{Observation of ultrastrong coupling and polariton formation.} \textbf{A} BLG transmission spectra at various displacement fields ($D$) under zero doping ($n=0$) measured at \SI{5}{K}. All spectra are referenced to the insulating state at $D=\SI{-0.36}{V/nm}$ and vertically offset for clarity. The cavity resonance frequency is indicated by an orange dashed line. A blue arrow points to the closing bandgap edge, while black arrows highlight the lower and upper polariton branches. \textbf{B} Contour plot representation of the spectra in panel A. The blue line represents the bandgap calculated using the model from ref.~\cite{mccann2013}, and the white dashed line marks the CNP position. \textbf{C} Calculated spectral function as a function of bandgap. The spectral function is referenced to the one of the bare cavity.}
    \label{fig:image4}
\end{figure}
\section*{Ultrastrong Coupling and Cavity-driven pairing}
The transmission spectra as a function of displacement field $D \neq 0$ for $n=0$ are shown in Figures \ref{fig:image4}A (line cuts) and \ref{fig:image4}B ($D$ vs frequency map). For decreasing fields, an absorption edge emerges and shifts from higher to lower frequencies (blue line in Fig.~\ref{fig:image4}A). This feature is consistent with the field-tunable band gap of BLG, measured here spectroscopically for the first time in this frequency range. The calculated field dependence following the self-consistent Hartree model for screening of ref.~\cite{mccann2013} is shown in Figure \ref{fig:image4}B (blue line), in excellent agreement with the measured edge.\\
As the gap approaches the cavity frequency with further decreasing fields, the spectrum undergoes dramatic changes. Two dispersive peaks develop: a strong, narrow peak below and a weak, broad peak above the cavity frequency. These modes correspond to the lower (LP) and upper (UP) polaritons, forming around a deepening spectral gap (the blue dip in Fig.~\ref{fig:image4}B), characteristic of a strongly coupled light-matter system.
The estimated coupling strength, defined by the ratio between the (effective) vacuum Rabi ($\Omega$) and the cavity ($\omega$) frequencies, is $\eta = \Omega/\omega\approx\SI{1}{THz}/\SI{2.3}{THz}=43\%$, which places our system well into the ultrastrong coupling regime ($\eta>10\%$ \cite{FriskKockum2019,forndiaz2019}). \\
In conventional ultrastrong coupling experiments, one typically deals with a single \emph{discrete} matter resonance—such as a cyclotron resonance, an exciton line or a bound-to-bound intersubband transition—that hybridizes with the cavity mode to produce two well-defined polariton branches \cite{deliberato2015,Duarte2025}. By contrast, in our system it is a \emph{continuum} of interband electron–hole transitions \cite{averkiev_light-matter_2007} that couples to the cavity field. In a continuum, the oscillator strength of each individual transition is usually much smaller, and the transitions have broader linewidths due to enhanced dissipation and momentum dispersion, which together drastically reduce the effective light–matter interaction. Yet, in our system, all states near the band edge contribute collectively, allowing the cavity to engage the entire continuum.
This observation hints at a fundamentally different mechanism: as predicted in ref.~\cite{khurgin2001}, the cavity field can act as an \emph{effective pairing interaction} that binds electron–hole pairs across a broad momentum range. This photon-mediated interaction reorganizes the continuum into a discrete state: the lower polariton (LP, black line in Fig~\ref{fig:image4}B) emerges as a bound-like state “pulled out” of the continuum (the white-hatched area below the gap and cavity lines in Fig.~\ref{fig:image4}B), while the upper polariton (UP, black line in Fig.~\ref{fig:image4}B) merges with the remaining high-energy transitions. Such a continuum-to-discrete reorganization departs significantly from standard ultrastrong coupling physics and illustrates the profound ability of vacuum electromagnetic fields to mediate attractive interactions in two-dimensional materials, paving the way for exploring photon-mediated pairing and emergent correlated phenomena.

\begin{figure}[h!]
    \centering
    \includegraphics[width=\textwidth]{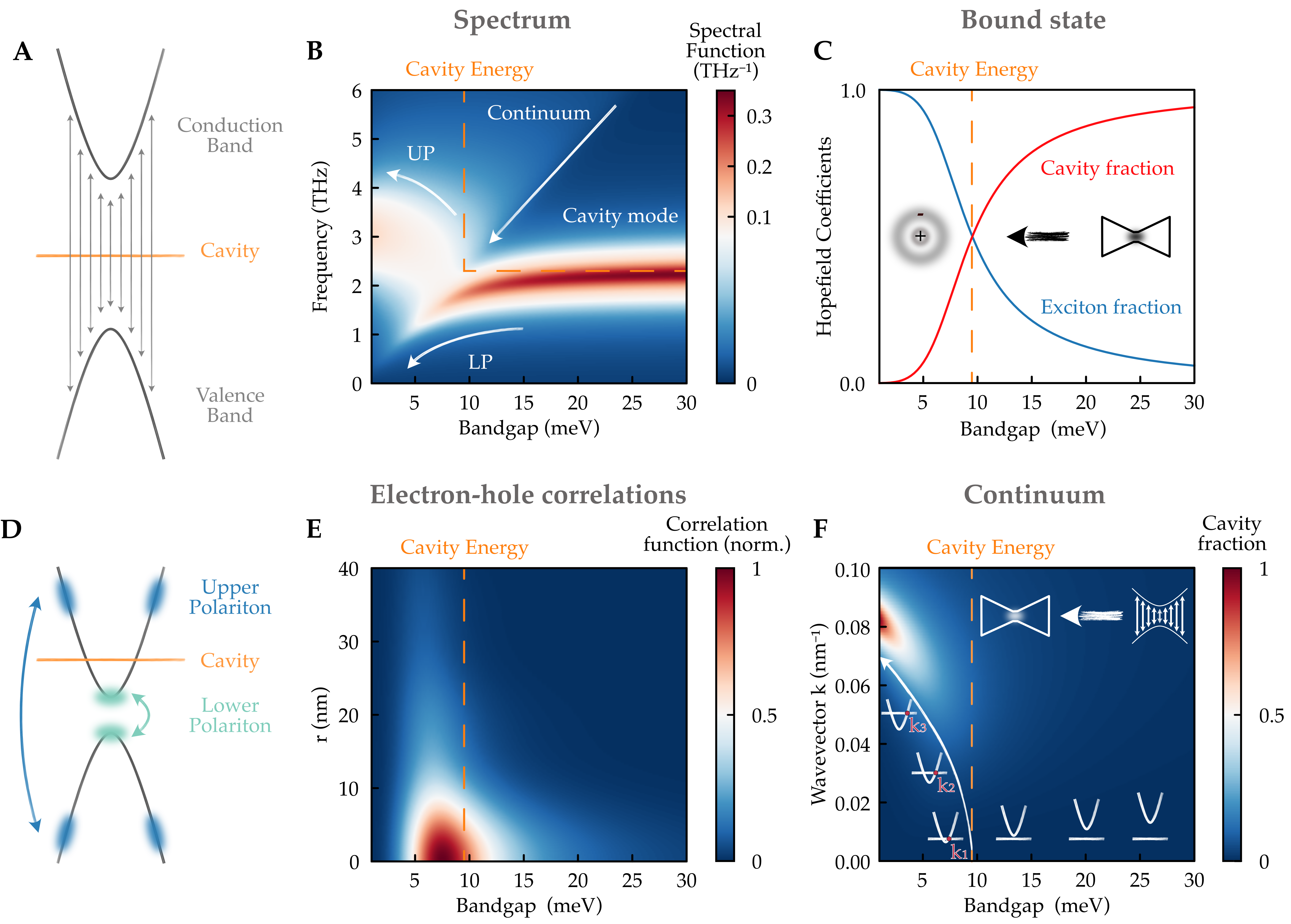}
    \caption{\textbf{Theory | cavity-induced bound state.} \textbf{A} Schematic representation of the model. A continuum of electron-hole excitations interacts with a single cavity mode. \textbf{B} Calculated spectral function for the ultrastrongly coupled system as a function of the bandgap. Note that only the (physical) spectrum for positive frequencies is displayed. For negative frequencies the spectrum exhibits a qualitatively similar behavior (see Supplementary Material). \textbf{C} Cavity and integrated matter content of the bound state as a function of gap energy. \textbf{D} Schematic of polariton formation. The lower polariton is primarily formed by states at the band edge whereas the upper polariton has its weight at higher electronic momenta. \textbf{E} Electron-hole correlation function $\gamma(r)$ of the bound state as a function of electron-hole distance $r$ and bandgap energy. The scale has been normalized to the maximum for clarity. \textbf{F} The cavity content of the continuum states is plotted as a function of bandgap and momentum $\hbar k$, where $\hbar$ is the reduced Planck constant and $k$ is the electronic wavevector. The white line marks the momentum at which the dispersionless cavity mode intersects the parabolic electron-hole dispersion.}
    \label{fig:image5}
\end{figure}

To support this interpretation, we have developed a minimal quantum mechanical model in which the two-dimensional conduction and valence bands are approximated by parabolic dispersions separated by a tunable gap. In this framework, each electronic momentum represents a potential electron–hole transition that couples to the cavity field with rate $g$ (see Fig. \ref{fig:image5}A). Importantly, by neglecting Coulomb interactions, any bound state formation can be attributed solely to the photon-mediated interaction. Operating in the ultrastrong coupling regime requires including counter-rotating (anti-resonant) terms \cite{ciuti_quantum_2005,Li2018}. This model reproduces the main experimental spectral features (Fig.~\ref{fig:image4}C) and offers deeper insight into the underlying physics. Our approach generalizes the theories developed for bound-to-continuum transitions \cite{Cortese_2019,Kumar_2022} to a fully continuum system. 
We first analyze the photonic spectral function (Fig.~\ref{fig:image5}B). For bandgaps much larger than the cavity frequency, the system consists of a discrete, bound cavity mode and a continuum of electronic transitions above the gap. As the bandgap decreases, the cavity mode hybridizes with the continuum, giving rise to two polariton branches. Focusing on the lowest-energy mode—which initially is predominantly photonic—we observe its gradual transformation into the lower polariton as the gap approaches the cavity frequency. This evolution is clearly reflected in the Hopfield coefficients (Fig. \ref{fig:image5}C), which reveal a shift from a predominantly cavity-dominated state to one that is largely matter-like. Importantly, as the hybridization with the continuum occurs, the emerging state retains the inherent bound character imparted by the cavity photons. Note that the matter component of the lower polariton is primarily composed of electron–hole transitions near the band edge, corresponding to low-momentum states (Fig. \ref{fig:image5}D).
To further probe the nature of this emergent bound state, we calculate the electron–hole correlation function as a function of the bandgap (Fig. \ref{fig:image5}E). As the bandgap approaches the cavity frequency, the correlations become significantly enhanced and more localized—closely resembling those of a conventional Coulomb-bound exciton. Notably, our experiment does not detect naturally occurring excitons even for bandgaps far above the cavity frequency  \cite{ju2017}. We attribute this absence to the inherently low binding energies and oscillator strength of excitons at the vanishing small bandgaps considered here—where the exciton Bohr radius can exceed tens of nanometers \cite{Li_exciton_2019,Henriques_exciton_2022,Sauer_exciton_2022,quintela20224} — and to strong Coulomb screening by the nearby gates, located approximately \SI{20}{nm} from the BLG layer.
Yet, we find that the effective photon-mediated interaction in our cavity is still able to bind electrons and holes.

Finally, we consider the upper polariton, whose matter component is predominantly formed by high-energy transitions involving large-momentum states (Fig. \ref{fig:image5}D). As the bandgap is reduced below the cavity frequency, these continuum states become progressively dressed by cavity photons, as illustrated in Fig. \ref{fig:image5}F. 
\section*{Temperature Dependence}
\begin{figure}[h!]
    \centering
    \includegraphics[width=\textwidth]{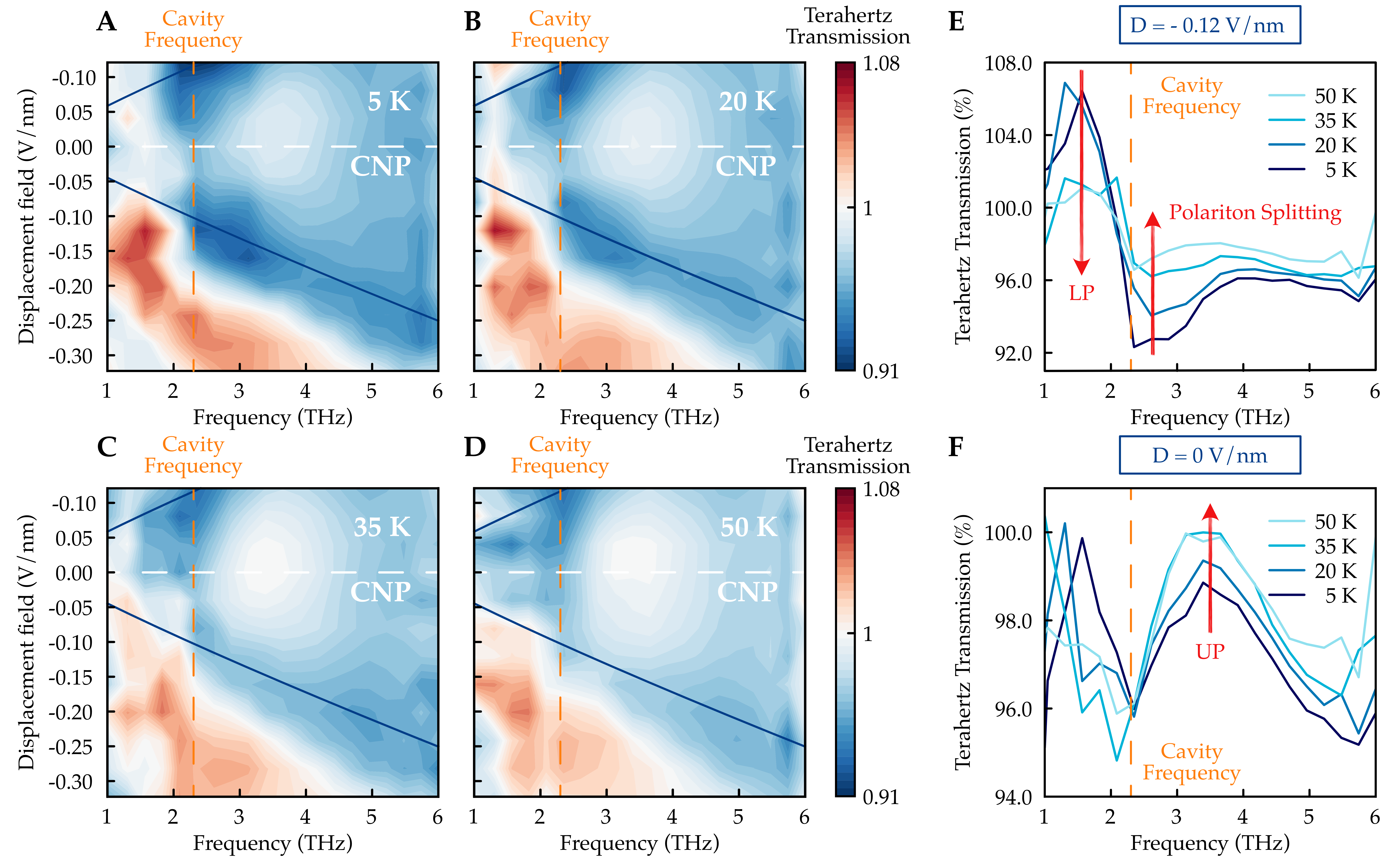}
    \caption{\textbf{Temperature dependence.} \textbf{A-D} Contour plots of the transmission spectra of BLG for different displacement fields $D$ and zero doping ($n=0$) measured at \SI{5}{K} (A), \SI{20}{K} (B) \SI{35}{K} (C) and \SI{50}{K} (D). All spectra are normalized to the transmission spectrum of the insulating phase obtained at $D=\SI{-0.38}{V/nm}$ at the respective temperature. The calculated bandgap is represented by a solid blue line, while a white dashed line indicates the CNP. \textbf{E, F} Line cuts extracted from the spectra presented in panels A-D at $D=\SI{0.12}{V/nm}$ (E) and at $D=\SI{0.00}{V/nm}$ (F).}
    \label{fig:image6}
\end{figure}

Figure \ref{fig:image6} shows the temperature dependence of the polariton spectra at 5 K (A), 20 K (B), 35 K (C), and 50 K (D). As the temperature increases, the contrast of the spectral features diminishes—most notably, the lower polariton and the spectral dip become less pronounced (Fig. \ref{fig:image6}E), while the upper polariton grows stronger (Fig. \ref{fig:image6}F). This behavior can be understood by noting that elevated temperatures enhance non-radiative decay processes and promote charges into the conduction band. As a result, states near the band edge are susceptible to stronger dissipation and decoherence—diminishing the intensity of the lower polariton—whereas higher-energy states are less affected (Fig. \ref{fig:image5}D), allowing the upper polariton to remain robust.  

Despite the increased dissipation, the key ultrastrong coupling features—such as the size of the polariton splitting—remain largely intact (see Fig. \ref{fig:image6} and Supplementary material). This observation supports the conclusion that the mechanism—photon-mediated pairing—is inherently robust against temperature. The splitting depends primarily on the oscillator strength and cavity confinement, intrinsic features of the light–matter interaction, rather than on thermal dissipation or decoherence. These results are in line with theoretical predictions and experimental studies demonstrating that the coupling strength, and hence the Rabi splitting, persists even at temperatures where the thermal energy exceeds the light–matter coupling energy, remaining largely temperature-independent \cite{deliberato2017, dini_microcavity_2003, anappara_giant_2007,geiser_2010, laitz_uncovering_2023}. This statement holds as long as the matter and cavity resonances do not qualitatively change their character as e.g. for conventional Coulomb-bound excitons when the thermal energy exceeds the exciton binding energy.
% Notably, this behavior is reminiscent of intersubband transitions in parabolic quantum wells, where strong coupling persists at temperatures well above the light–matter coupling strength due to the equidistant energy levels effectively counteracting thermal depletion of the initial state \cite{geiser_2010}. 

% The temperature-dependent measurements reveal a distinctive aspect of our system. While previous studies in doped quantum wells have focused on bound-to-continuum transitions \cite{cortese_excitons_2021}, our work demonstrates pairing emerging solely from continuum-to-continuum transitions. This distinction is significant for two reasons: it more faithfully reflects the interactions present in real many-body systems, and it results in a pairing mechanism with markedly reduced sensitivity to temperature, as thermal excitation promotes electrons from the valence to the conduction band without substantially depleting the initial state.

\section*{Discussion}
We have demonstrated that THz cavities serve as powerful tools for generating attractive interactions in quantum materials. To validate this phenomenon, we developed an experimental platform specifically designed for the far-infrared spectroscopy of microstructured WdV heterostructures. Our measurements reveal two distinct regimes: when the cavity is off-resonance, we probe the intrinsic THz bandgap of BLG, providing an unparalleled window into the material's optical properties; conversely, tuning the system into resonance triggers hybridization—a hallmark of ultrastrong coupling—where the spectroscopic signature reflects the coupled light–matter system, thereby enabling the observation of cavity-induced bound states. Our platform addresses the limitations of conventional THz spectroscopic techniques for micro-structured devices, which often suffer from bandwidth constraints or incompatibility with dual gating. While on-chip terahertz techniques have proven highly effective in probing the ultra-low energy electrodynamics of 2D materials \cite{gallagher2019,mciver2020,potts2023,zhao2023,kipp2024}, they face inherent bandwidth limitations \cite{grischkowsky2000}, typically not exceeding 1.5 THz or 6 meV, as well as a lack of tunability. Notably, much of the physics of quantum \cite{Basov2011} and VdW materials \cite{novoselov_2d_2016} is expected or has been observed at energies of several to tens of meV. This energy range encompasses collective modes such as phonons \cite{Haastrup2018}, Mott gaps and electronic bands in Moiré systems \cite{yang2022}, binding energies \cite{Ma2021} as well as Higgs and Bardasis-Schrieffer modes in excitonic insulators \cite{sun2020,xue2020}, and gaps and excitations of Wigner crystals \cite{brem2022}, among others. With a bandwidth exceeding 20 meV, our technique outperforms existing methods and bridges the gap between on-chip terahertz and photoconductivity measurements \cite{ju2017,yang2022,Bandurin2022,Ma2023} while providing access to time resolution, spatial resolution ($\ll $ wavelength) as well as the full material optical properties through amplitude and phase measurements of the transmitted field. In addition, our method offers a significant advantage over conventional THz scanning-probe techniques, which, despite achieving spatial resolutions of tens to hundreds of nanometers, struggle with full dual gating and require extensive calibration and detailed understanding of the probe-sample interaction to extract reliable spectroscopic information \cite{guo_tip2024}. Lastly, our approach is versatile, being adaptable to various generation techniques and cavity designs. This flexibility makes it scalable across a wide range of frequency domains, from the terahertz to the mid-infrared. 

Looking ahead, integrating our technique with dilution refrigerators capable of reaching millikelvin base temperatures represents the next frontier. By delivering low-power, low-photon-energy THz radiation directly to the sample, our method circumvents the heating effects typically encountered with optical laser pulses in on-chip THz setups. This advance will enable broadband far-infrared spectroscopy of 2D materials under ultra-low-temperature conditions, thereby opening the door to investigating delicate quantum phenomena—including superconductivity \cite{Cao2018}. Given that bilayer graphene exhibits a propensity for superconductivity at large displacement fields and millikelvin temperatures \cite{zhou2022}, our platform holds significant promise for exploring cavity-assisted superconducting phase transitions \cite{lu_cavity-enhanced_2024}, paving the way for engineering and control of correlated phases via tailored light–matter interactions. 

Lastly, our findings raise fundamental questions about the cavity‑controlled ground state. The appearance of excitonic features at finite bandgaps prompts us to ask whether the cavity‑induced bound state survives once the gap fully closes. The observation of an upper polariton without any applied displacement field further challenges our understanding of the vacuum‑dressed ground state. Future experiments that directly probe ground‑state coherence could uncover new ordered phases—such as cavity‑driven excitonic insulators\cite{Ma2021}. In general, the highly localized, non‑uniform cavity fields used here provide a powerful testbed for rigorously exploring cavity QED phenomena\cite{li_observation_2018,kim_supperradiant_2025} and scrutinising predictions—including the debated no‑go theorems governing vacuum‑field–driven phase transitions\cite{mazza2019,andolina2019,andolina2020,ashida2020,curtis2023,liu_magnetopolariton_2014}.

By combining strong sub‑wavelength confinement with gate‑tunable van der Waals heterostructures, our approach opens an exciting avenue for engineering many‑body phenomena via electromagnetic vacuum fields and paves the way for deeper investigations into cavity‑induced phase transitions and the fundamental nature of light–matter interactions in quantum materials.
\section*{Acknowledgements}
The authors thank E. J\"ochl for discussions on the lens assembly and sample processing. F.H. acknowledges discussions on sample fabrication with the Ensslin group (J. Gerber, M. Niese and M. Ruckriegel), support from the FIRST cleanroom staff and M. Baer for the design of the sample holder. T.F.N. thanks M. Buzzi and M. Schiro for fruitful discussions. The BNA THz generation crystals were provided by Swiss Terahertz LLC. T.F.N. was supported by the ETH Postdoctoral Fellowship Program. F.H. and A.I. acknowledge support from the Swiss National Science Foundation (SNF) (Grant number  200020 207520). J.F and G.S. acknowledge funding by the Swiss National Science Foundation (SNF) (Grant number 10000397). H.S.A. acknowledges support from the Swiss Government Excellence Scholarship. I.K. acknowledges the financial assistance of the Rothschild Post-Doctoral Fellowship from Yad HaNadiv, the Helen Diller and the Viterbi Post-Doctoral Fellowships from Technion.

%somewhat weaker...
%Lastly, our findings underscore the exciting potential of cavity-mediated interactions for engineering pairing in correlated electron systems. Over the past decade, theory has advanced two complementary paradigms. First, direct photon-mediated pairing—where virtual photon exchange yields a current–current attraction—can drive exotic finite-momentum (pair-density-wave) superconductivity, albeit with modest transition temperatures under current coupling conditions \cite{schlawin2019}. 
% Macroscopically occupied polariton states offer a similar but distinct cavity-induced pairing mechanism, also predicted to lead to superconductivity \cite{cotlet_2016}. Second, indirect cavity-enhanced pairing leverages the cavity to bolster traditional mechanisms (such as phonon-mediated interactions), offering a route to boost superconductivity even in systems already predisposed to pairing \cite{lu_cavity-enhanced_2024}.
%Notably, superconductivity in two-dimensional materials often exists in the dilute regime, where carrier densities are low. When combined with the strong field confinement achievable in our setup, this opens the intriguing possibility of reaching an ultrastrong coupling regime at the level of individual electrons or even single Cooper pairs. Such a regime might pave the way for designing and controlling superconducting phases through tailored light–matter interactions.

\bibliographystyle{unsrt}
\bibliography{bibliography}
\end{document}